# Soccer: a quantitative analysis of team resilience and the miracle of Bern

Ralph Stömmer*

(Working paper, version 2, November 2, 2026)

## Abstract

Resilience is the ability to positively respond to adversity. It has been studied in psychology for several decades, with focus on how individuals overcome traumata or cope with setbacks and obstacles in their professional career. Research on resilience in the sport context is rather new. Activities are based on insights that in highly competitive environments, tiny effects tip the scales. A key question of measuring resilience in sports is what parameters to measure. Here a novel concept is proposed to measure the resilience of soccer teams. The relative frequency of matches is determined, where a soccer team, which is initially trailing by 2 goals, finally succeeds to win the match or at least to reach a draw. The analysis is applied to the last 59 seasons of the German premier soccer league Bundesliga. The empirical data are compared with a theoretical model derived from Poisson distributions. It is shown how leading teams in the premier soccer league differ from the average with respect to resilience, which provides further insights into the hidden secrets of top soccer teams.

---

* Private researcher, e-mail: ralph.stoemmer@trajectorix.de

## 1. Introduction

For a wide variety of domains spanning from working life over politics to sports, literature is highlighting the potential gains of positive thinking and resilience. A significant correlation was found between optimism and athletic performance of soccer players [1]. In a frequently cited study dating back to the year 1990, Seligman et al. [2] found a relationship between the athletes explanatory style (optimistic versus pessimistic) and performance in swimming competitions. Although not explicitly mentioned and probably not of primary focus over 30 years ago, resilience was implicitly tested in Seligman´s work. When the athletes performance was benchmarked and their explanatory style investigated, they were given falsely negative swimming times to experimentally simulate unexpected defeat before they went to their next swim. Initial and unexpected setbacks are crucial for the concept of resilience. Individuals and teams don´t face their challenges under benevolent or under neutral conditions, they do receive stress and obstacles instead. Resilience includes the ability to overcome barriers, to cope with negative conditions, and to respond positively to failures and to adverse events. To operationalize the term, resilience implies two conditions: (i) adversity and (ii) positive adaption despite this adversity [3]. Although resilience and positive thinking seem to be correlated, there is an important distinction. Positive thinking is an attitude mainly located in the realm of judgement. It is primarily accessed by psychological inquiry, with the drawback that test persons might align answers towards desired responses and overestimate their own abilities [4]. Resilience relates to experience and strength of nerve, it does reach beyond positive thinking into the realm of decision making and acting. It can be accessed by benchmarking objective results, thereby dispensing with potentially biased self-assessments. For sports, a key question of quantifying resilience is what parameters or what results to measure [5].

Here a novel concept is proposed to measure the resilience of soccer teams. The relative frequency of matches is determined, where a soccer team, which is initially trailing by 2 goals, finally succeeds to win the match or at least to reach a draw. A 0:2 leeway is defined as the initial setback for the soccer team, because it is experienced as a considerable nuisance. A potential discouragement effect relates to the expectation value E of goals per match, which is as low as $E \approx 3$ for the German premier soccer league [6]. Being 2 goals behind in a low-scoring sports like soccer might constitute a preliminary decision for the final loss. The chances for catching up are low, as calculated probabilities from theory and relative frequencies from empirical data illustrate in the results section of this work.

Depending on the team attitude, the prospect of losing the match can be a source of despair which triggers a performance drop, as described for the swimmer example [2]. Alternatively, trailing might set free positive responses to change the game, where the team finally succeeds to win the match or at least to balance the score by a draw. The process in which a soccer team initially suffers a leeway, but turns the final score to its favor by a win or a draw, is interpreted as an instance of resilience.

A prominent example for an initial 0:2 leeway and a final 3:2 win is the match of the soccer finalists Germany and Hungary in the 1954 FIFA world cup in Bern, Switzerland. The contest appeared done, because the two teams had already met in the group phase before with Hungary triumphing 8:3. Hungary, unbeaten in four years, looked set to claim its very first world cup title. But the final match went unexpected: Germany was 0:2 behind first, but then managed to change the odds and surprisingly won the match 3:2. This incidence is generally known as the “miracle of Bern”, which evoked a wave of euphoria throughout Germany [7]. The following analysis will disclose the probabilities of such outcomes.

## 2. Sources of data

The analysis is based on 17879 matches with 54679 goals from 59 seasons of the German premier soccer league Bundesliga, starting with season 1963/64, and including the season 2021/22 up to the 21$^{st}$ match day. Data are taken from the online databases www.bundesliga-statistik.de and www.forum.vmlogic.net, which complement each other regarding the results for each season and the summaries provided in the all-time table. The Bundesliga ranking does consist of 18 teams, with changing composition over the years, because teams ascend from the lower league into the Bundesliga if they qualify, or they relegate to lower leagues in case of poor performance. Since start of the Bundesliga in the season 1963/64, in total 56 different soccer teams have been part of the premier soccer league. Each of the 18 soccer teams of the league plays against any of the other 17 teams twice a season, one home match and one away match to balance the home advantage. In each season, in total 306 matches are taking place, except for the initial seasons 1963/64 and 1964/65, where the Bundesliga started with 16 teams accounting for just 240 matches each season, and the season 1991/92, where the Bundesliga consisted of 20 teams accounting for 380 matches.

## 3. Theoretical model

### 3.1. Conversion of Poisson distributions into a Poisson an a binomial distribution

Scoring goals in soccer matches is described as a stochastic process. It can be understood as a Bernoulli experiment, where the occurrence of each goal has a fixed probability, and events do not influence each other. To a good approximation, the distribution of the number of goals in a match between two teams A and B is taken as the product of two independent Poisson distributions [8, 9, 10, 11, 12]. The probability p(k,l) for the match result is given by

$$p(k,l) = \frac{E_A^k}{k!} e^{-E_A} \frac{E_B^l}{l!} e^{-E_B} . \qquad (1)$$

k denotes the number of goals scored by team A, l denotes the number of goals scored by team B. Match results are generally provided as the integer relation k:l. $E_A$ and $E_B$ denote the expectation values of goals per match achieved by team A and team B, respectively. In many cases, the expectation value E of goals per match is more reliably determined from a large number of matches of a whole soccer league. With $k + l = m$ and $E_A + E_B = E$, equation 1 is provided in a form which is more convenient for further analysis:

$$p(k,l) = \frac{m!}{k!\,l!} \left(\frac{E_A}{E}\right)^k \left(\frac{E_B}{E}\right)^l \frac{E^m}{m!} e^{-E} . \qquad (2)$$

p(k,l) is the product of a binomial distribution p(k,l/m) and a Poisson distribution p(m). p(m) provides the probability for in total m goals scored by two teams in matches where E goals are to be expected [6]:

$$p(m) = \frac{E^m}{m!} e^{-E} . \qquad (3)$$

The histogram in figure 1 demonstrates that the relative frequencies derived from empirical data are well described by a Poisson distribution. The relative frequencies of goals scored are based on 17879 matches of the soccer league Bundesliga accomplished in 59 seasons from the starting season 1963/64 up to the 21st match day of the season 2021/22. The theoretical Poisson distribution p(m) in equation 3 is inserted into figure 1 as a red dashed line, with expectation value $E = 3.1$. The data and the theoretical model are plotted versus the number m of goals scored and provided on a logarithmic scale, which allows to visibly reveal where data and theory fit to each other, and where they deviate from each other. For $m = 0$, which corresponds to 0:0 draws, data exhibit a higher value compared to theory. This deviation is dealt with by Rue et al. [13], who observed that the amount of draws in soccer matches is slightly larger than expected with independent Poisson distributions. For high scores $m > 9$, relative frequencies exceed theoretical values, too. This phenomenon is described in the context of heavy tail or fat tail distributions, which are observed in a wide range of domains from finance over physics to earthquake data [ 14, 15, 16]. Heavy tails indicate that extremes are more common than theory does anticipate. Events are not fully independent, as assumed for simplified Bernoulli trials, but are slightly correlated. For soccer, heavy tails at high scores can be illustrated assuming two competing teams with extremely diverging team strengths. If the stronger team A is leading with 6:0, it is obvious that the weaker team B is probably discouraged completely. Team A might be affirmed by the extraordinary achievement to score additional goals. Theoretical models can be modified to include components of self-affirmation, which consider correlations between subsequent scoring events and describe the heavy tails of distributions surprisingly well [17]. For the further analysis in this work, fine tuning of the present model to address weak correlations is not required. First of all, high scores in soccer are comparably rare events, as the low probabilities in figure 1 demonstrate. Secondly, in the goal range of primary interest covering the values $m = 2 \ldots 5$, empirical data and theory match reasonably well.

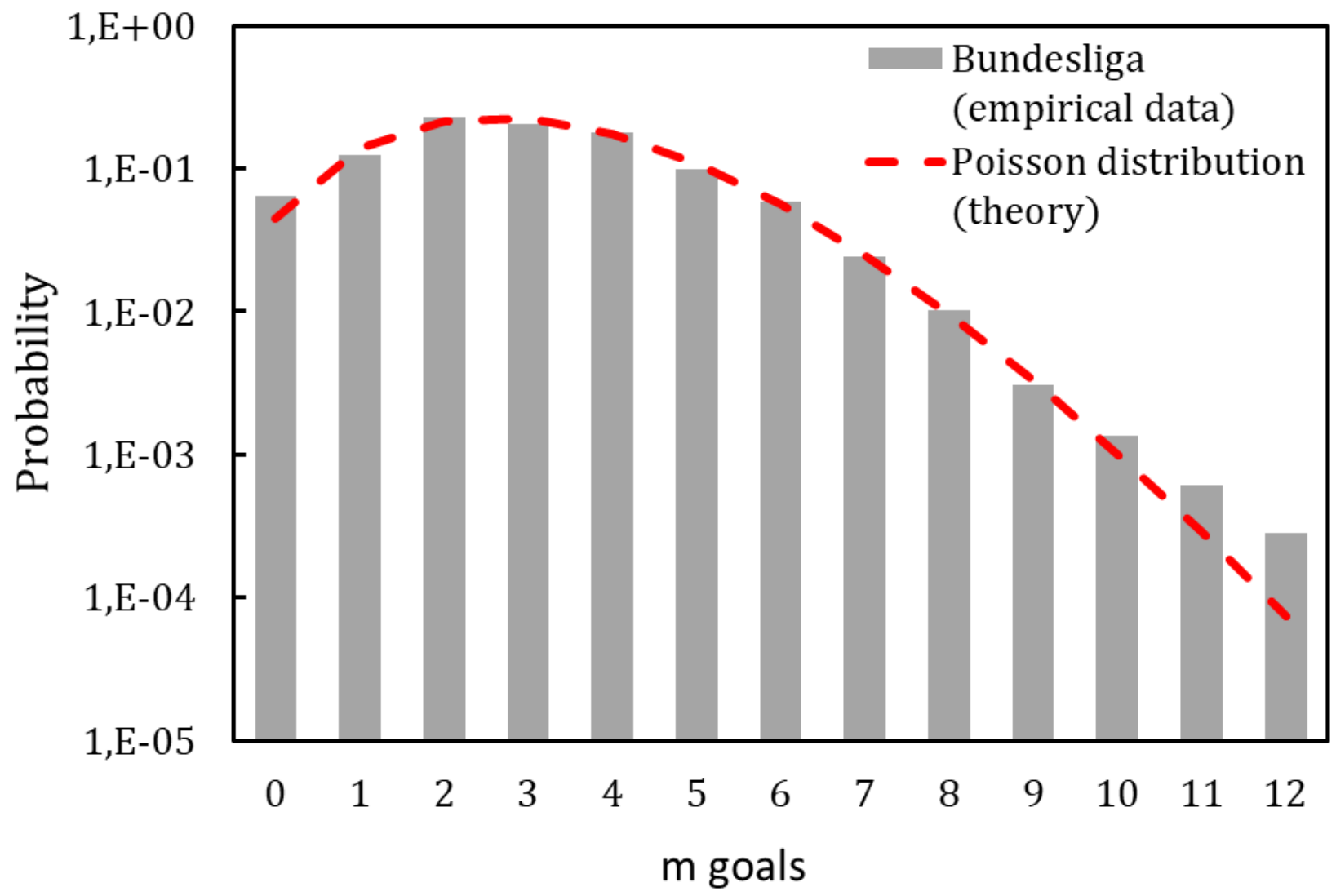

**Figure 1**: Histogram with relative frequencies of goals scored per match based on empirical data from the Bundesliga, plotted versus number of goals m, and corresponding theoretical Poisson distribution with expectation value E=3.1.

The binomial distribution p(k,l/m), which precedes the Poisson distribution p(m) in equation 2, provides the probability for a k:l result under the condition that in total m goals are scored:

$$p(k,l/m) = \frac{m!}{k!\,l!}\left(\frac{E_A}{E}\right)^k \left(\frac{E_B}{E}\right)^l \ . \tag{4}$$

The two fractions in equation 4 can be rewritten as $E_A/E = p_A$ and $E_B/E = p_B$. $p_A$ denotes the goal scoring probability of team A, and $p_B$ denotes the goal scoring probability of team B. With $p_B = 1 - p_A$ and $l = m - k$ we obtain

$$p(k,l/m) = \frac{m!}{k!(m-k)!} p_A^k \,(1-p_A)^{m-k} \ . \tag{5}$$

### 3.2. Derivation of a quantitative measure for resilience from the binomial distribution

The goal scoring probability $p_A$ of a team is a measure for its relative team strength. Without any prior knowledge, $p_A$ can be set to 50% to account for equal strength when both teams meet on neutral ground. In order to consider the home advantage, one can increase $p_A$ to values > 50% if team A is the home team, or $p_A$ < 50% if team B is the home team. In case team A is trailing by 2 goals, which implies that team B has scored 2 goals in a row first, equation 5 yields for the probability pT(0,2) of a 0:2 leeway:

$$pT(0,2) = (1-p_A)^2 \,. \tag{6}$$

The probability to achieve a 2:2 draw after trailing by 2 goals calculates to

$$p(\text{draw}) = pT(0,2)p_A^2 \ . \tag{7}$$

The probability to win 3:2 after trailing by 2 goals calculates to

$$p(\text{win}) = pT(0,2)p_A^3 \ . \tag{8}$$

Equations 7 and 8 do not contain the permutation term on the right hand side in equation 5, because the goals are achieved in a fixed sequence. Changing the odds and making up leeway after trailing by 2 goals is given by p(win or draw) = p(win) + p(draw), which yields

$$p(\text{win or daw}) = pT(0,2)p_A^2(1+p_A) \ . \tag{9}$$

When the goal scoring probability $p_A$ remains constant throughout the match, p(win or draw) ≤ 9.5%. p(win or draw) itself is no indication for resilience, it just provides the probability for changing the odds for a given $p_A$. Resilience takes effect when $p_A$ increases after suffering a 0:2 leeway. As a consequence, the probability for a win or a draw increases beyond the value p(win or draw) initially calculated.

For illustration, let us assume two teams starting the match with equal strengths. $p_A = 50\%$ put into equation 6 yields pT(0,2) = 25%, and equation 9 yields p(win or draw) = 9.4%. When team A suddenly enhances its performance to $p_A = 80\%$ after trailing by 2 goals with pT(0,2) = 25%, the probability for a win or a draw would increase from 9.4% to 29%. When statistical data are collected, resilience should reveal itself by the positive offset between the empirical relative frequency h(win or draw), which might contain resilience effects, and the theoretical probability p(win or draw) calculated with constant $p_A$ without resilience effects:

$$\text{delta(resilience)} = \text{h(win or draw)} - \text{p(win or draw)}\ . \tag{10}$$

The theoretical probability p(win or draw) is applied as the soccer team´s expected base value for achieving a win or a draw after trailing by 2 goals. If resilience is absent in empirical data, h(win or draw) does match with p(win or draw). It is emphasized that the quantities in equation 10 are statistical in nature. If resilience in a soccer team is present on occasion, but cancels out for the number of matches investigated, delta(resilience) = 0. This cannot be distinguished from a soccer team where resilience is absent all along. Equation 10 can lead to delta(resilience) < 0, when the soccer team fails to meet or fails to exceed the expected base value p(win or draw). Extending the definition for resilience provided in the introduction [3], a negative value would correspond to a negative adaption to adversity. It corresponds to a reduction of the goal scoring probability $p_A$ after trailing by 2 goals. Relevant literature is short on negative resilience effects, for the purpose in this work it can be circumscribed at best along terms like vulnerability and discouragement.

## 4. Results

### 4.1. Magnitude and compensation of the home advantage

The home advantage reveals itself as an asymmetry in the relative frequencies of goals scored at home and at away matches, it has a considerable effect onto results. For investigating resilience, the home advantage needs to be quantified first and compensated for further analysis. Figure 2 shows the histograms derived from empirical data from 17879 matches with 54679 goals, consisting of 32976 home goals and 21703 away goals, which does yield the expectation value of goals per match E = 3.1. The relative frequencies of goals, which for the large number of goals achieved approach the probability of goals p_home for the home team and p_away for the away team, are plotted as colored dashed lines versus season periods, starting with season period 1963/64 – 1971/72. Over all seasons, the mean values are p_home = 60% and p_away = 40%. Without home advantage, the probability for each team to score a goal would equal p = 50%, provided as a red dashed horizontal line. The colored columns denote the relative frequencies from empirical data for trailing by one goal at a home match hT_home(0,1) and at an away match hT_away(0,1). The red columns provide the relative frequency hT(0,1) for conceding the first goal, compensated for the home advantage, as if the competing teams would meet on neutral ground. For clarity, the data are aggregated into periods of 10 years, with the first period containing 9 years only. Grouping the data into decades accounts for the findings of Heuer et al. [18], who determined two time regimes in a soccer club: in a short regime on time scales of approximately 2 years, team fitness doesn´t change much, it is biased mainly by short-term measures like tactics or team composition. The long regime occurs on time scales of decades (20 – 30 years), where fundamental properties of the team, the soccer club and the environment affect the performance, like the economic situation, changing strategies in the soccer league, overall professionalism and sportiness levels of athletes (just envision today´s constitution of soccer players and compare it with that of soccer players in 1982).

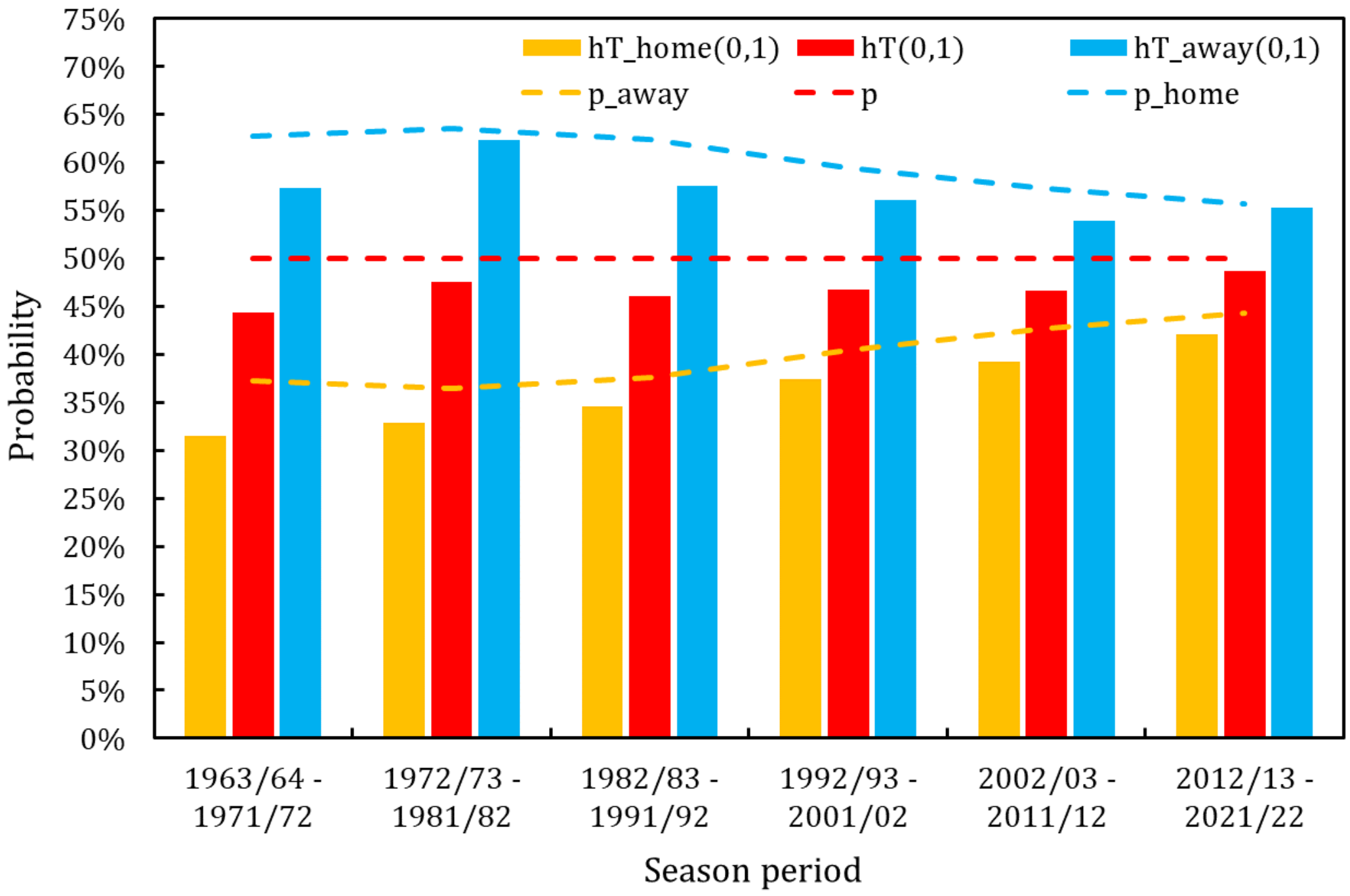

**Figure 2**: Histograms for trailing by one goal derived from empirical data from the Bundesliga, plotted versus season periods. hT_home(0,1) and hT_away(0,1) denote the relative frequencies for a 0:1 leeway at a home match and an away match, respectively. hT(0,1) is the relative frequency for a 0:1 leeway compensated for the home advantage. The dashed lines for p_home and p_away denote the probabilities to score a a goal at a home match and an away match. p is the probability for scoring a goal compensated for the home advantage.

Figure 2 shows that p_home and p_away change over time. For first estimations it can be assumed that the home team is twice as strong as the away team, p_home = 2 p_away [6]. With the condition p_home + p_away = 1, it follows that p_home = 2/3 and p_away = 1/3, which indeed was the case in the early decades of the Bundesliga. Both probabilities slowly approach p = 50% with a rate of -2% and +2% per decade for p_home and p_away, respectively. Extrapolated with a rate of -2%, the home advantage is going to vanish in the season period 2032/33 – 2041/42. The underlying mechanisms for the home advantage are still not well understood. Among other causes, the review from Pollard [19] contains the territoriality model, which would explain the decrease observed over the years. Humans are known to respond to a real or to a perceived threat of their home territory with increased performance, which was confirmed for soccer players, too [20]. This effect should diminish when the emotional ties to the territory loosen, as one would expect for professional soccer players, who change club affiliation based on strategic and economic reasons just like modern mercenaries. In this regard, the diminishing home advantage is the inevitable outcome of a thoroughly professionalized soccer league.

hT_home(0,1) and hT_away(0,1) do approach p_away and p_home, respectively. This is no coincidence, because the relative frequency for the home team to concede the first goal should come close to the probability p_away for the away team to score the first goal, and vice versa. The small offset is attributed to the fact that a few percent of all soccer matches end with no goal at all, p(0,0) = 4.5% according to equation 2.

### 4.2. Resilience in the German premier soccer league Bundesliga

Whereas a 0:1 leeway does produce some discomfort for the team, a 0:2 leeway is experienced as a considerable nuisance. Figure 3 compares the relative frequencies derived from empirical data for the home team hT_home(0,2) and the away team hT_away(0,2) for trailing by 2 goals, plotted versus season periods in a similar manner to figure 2. The theoretical probabilities pT_home(0,2) and pT_away(0,2) are calculated according to equation 6 and inserted as dashed lines. Again assuming that the home advantage would not exist, the histograms contain the relative frequency hT(0,2) and the corresponding probability pT(0,2) for a 0:2 leeway. A long term trend of hT_home(0,2) and hT_away(0:2) towards converging at ≈25% can be observed.

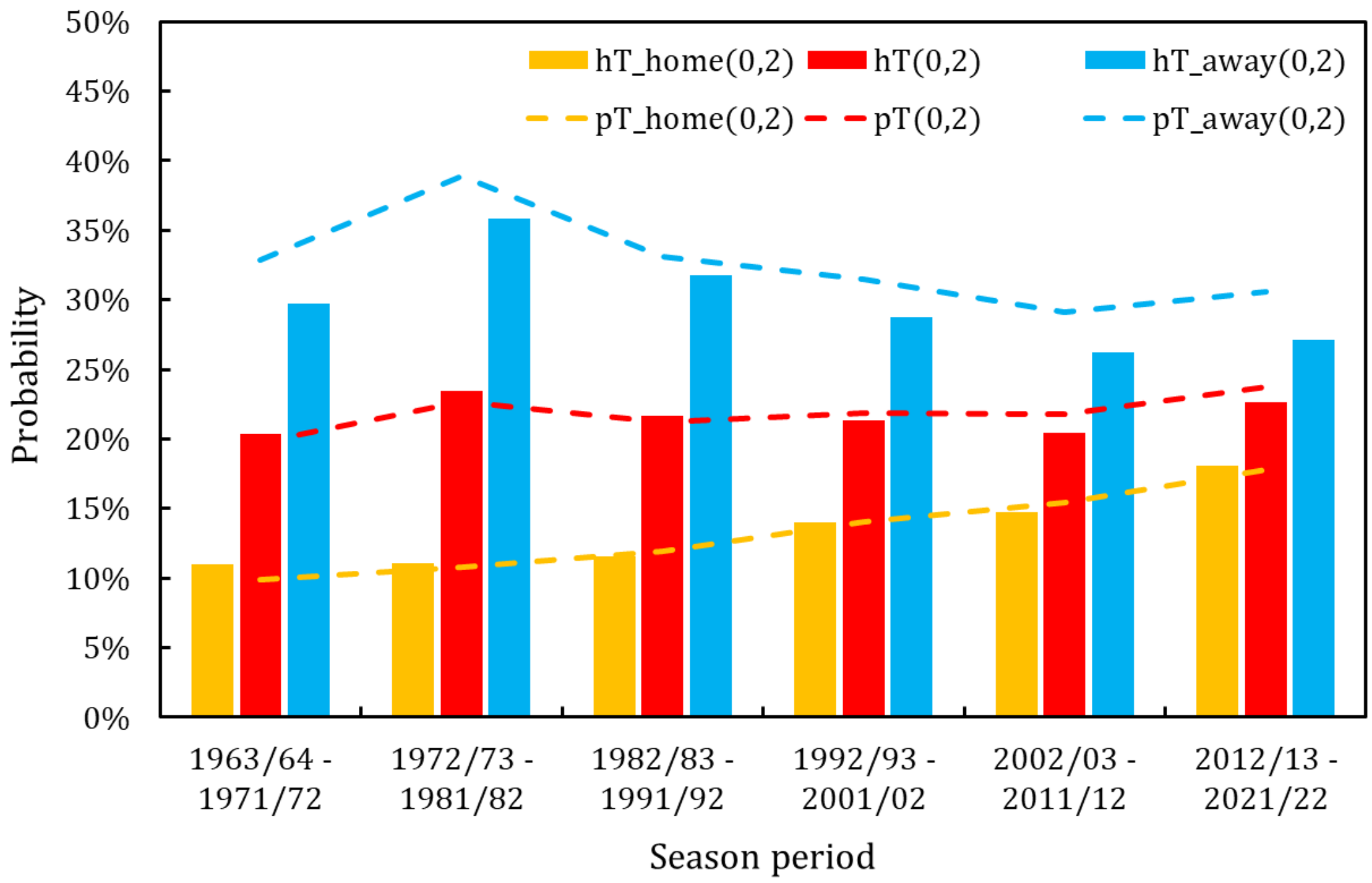

**Figure 3**: Histograms for trailing by 2 goals derived from empirical data from the Bundesliga, plotted versus season periods. hT_home(0,2) and hT_away(0,2) denote the relative frequencies for a 0:2 leeway at a home match and an away match, respectively. hT(0,2) is the relative frequency for a 0:2 leeway compensated for the home advantage. The dashed lines for pT_home(0,2) and pT_away(0,2) denote the calculated probabilities for a 0:2 leeway at a home match and an away match. pT(0,2) is the probability for a 0:2 leeway compensated for the home advantage.

For the further evaluation of resilience, relative frequencies hT(0,2) and probabilities pT(0,2) are employed where the home advantage is compensated. Inserting $p = 50\%$ from figure 2 and the average value $pT(0,2) = 21.8\%$ from figure 3 into equations 7 and 8 yields $p(draw) = 5.5\%$ and $p(win) = 2.7\%$. The overall probability to prevent a loss calculates to $p(win\ or\ draw) = 2.7\% + 5.5\% = 8.2\%$. The low probabilities again confirm why a 0:2 leeway is experienced as a setback for the team. It is challenging for the trailing team to score 2 goals in sequence to reach a draw or 3 goals in sequence to finally win the game. A 0:2 leeway requires the team to activate considerable resources and strength of nerve to enhance performance and to change the match to its favor.

Figure 4 shows the histograms with empirical data from the Bundesliga over all seasons under the condition that the teams suffered a preceding 0:2 leeway. h(win or draw) denotes the relative frequency to win the match or to achieve a draw, compensated for the home advantage. The relative frequencies h_home(win or draw) for the home team and h_away(win or draw) for the away team are inserted for reference. The dashed line denotes the theoretical probability p(win or draw) calculated with equation 9.

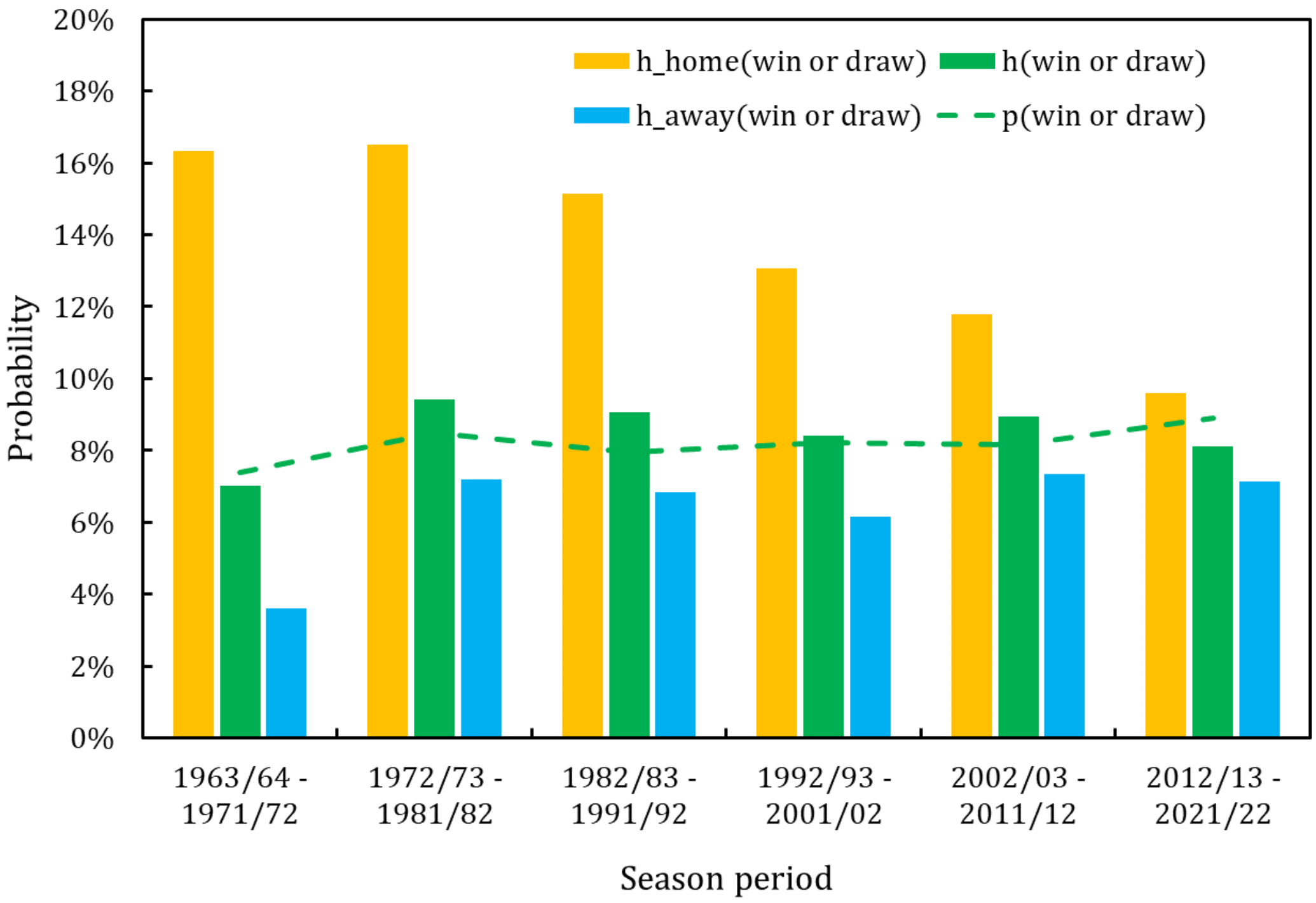

**Figure 4**: Histograms for winning the match or reaching a draw after having suffered a 0:2 leeway, derived from empirical data from the Bundesliga and plotted versus season periods. The relative frequency h(win or draw) is compensated for the home advantage. h_home(win or draw) for the home team and h_away(win or draw) for the away team are inserted for reference. The dashed line denotes the theoretical probability p(win or draw).

h(win or draw) as well as the corresponding p(win or draw) in figure 4 don´t display a clear long term trend, they are just framed by h_home(win or draw) and h_away(win or draw), which converge towards ≈9.4%. This can be explained as follows: the home advantage is slowly fading away, so chances are distributed equally between competing teams with $p_A$ = 50%. Inserting $p_A$ = 50% into equation 9 yields p(win or draw) = 9.4%. Surprisingly, relative frequencies from empirical data and theoretical probabilities do match closely in figure 4, which is a result one wouldn´t expect. The maximum deviation between theory and data is 1.1% in season period 1982/83 – 1991/92. If a team manages to win the match after a 0:2 leeway, it can win 3:2, but higher wins like 4:2 or 4:3 are possible, too. With respect to the Poisson distribution in equation 3 and probability values provided in figure 1, higher scores than 3:2 are quite rare events with just minor contributions.

Figures 5, 6a and 6b contain the all-time relative frequencies hT(0,2) and h(win or draw) of the whole soccer league Bundesliga and that of individual soccer clubs, which are part of the top 20 in the all-time table of the league. Up to now, 56 different soccer clubs have been part of the league for various periods from the starting season 1963/64 up to the current season

2021/22 (status 21st match day). The data for the whole league are aggregated in figure 5, the data for the top 20 clubs are provided in an equal manner, sorted from left to right in figures 6a and 6b according to their current ranking in the all-time table. The ranks are attached to the club names below the horizontal axis.

The majority of the top 10 clubs in figure 6a have been in the league for at least 50 years, except Bayer 04 Leverkusen, which advanced into the Bundesliga first time in the season 1979/80. In the current season 2021/22, 7 out of the top 10 clubs of the all-time table are still among the 18 active teams in the Bundesliga. Hamburger SV relegated into the lower league after season 2017/18, as well as Werder Bremen and FC Schalke 04, which relegated into the lower league after the last season 2020/21. For the clubs on the succeeding ranks 11 to 20 in figure 6b, the situation is vice versa. Only 3 out of 10 clubs are still among the 18 active teams in the Bundesliga, namely Hertha BSC, VfL Bochum, and VfL Wolfsburg. Except for VfL Wolfsburg, which ascended into the Bundesliga in season 1997/98 and permanently has been staying there, the presence of the other teams has been discontinuous, with periods of relegating into lower leagues, and ascending back into the Bundesliga on occasion.

The data for the entire league in figure 5 represent the values of all soccer clubs, with hT(0,2) = 21.7% denoting the relative frequency for trailing by 2 goals over all seasons (split into season periods, see figure 3). h(win or draw) = 8.6% is the relative frequency for turning the match into a win or a draw after an initial 0:2 leeway (split into season periods, see figure 4). The theoretical probability p(win or draw) is located as dashed column alongside the empirical relative frequency h(win or draw). According to equation 10, the difference between h(win or draw) and p(win or draw) is interpreted as a resilience effect, it is provided as yellow column delta(resilience) alongside p(win or draw). The histograms in figures 5, 6a and 6b should be understood as follows: the relative frequency for a soccer team to trail by 2 goals in a match is hT(0,2). From these matches with a 0:2 leeway, the team succeeds to change the odds to its favor with a relative frequency h(win or draw). h(win or draw) should match with the theoretical probability p(win or draw) in case the probability $p_A$ to score a goal does not change after having conceded 2 goals. This indeed is the case for the whole league in figure 5, with a minor deviation of 0.4%, which is attributed to the aforementioned rare events of winning matches with higher than 3:2 scores. As noted in the previous section 3, the quantities are statistical in nature. The model proposed in equation 10 for measuring resilience is not able to distinguish if resilience is present, but cancels out to zero with all matches investigated, or if resilience is absent all along. Figure 5 demonstrates for the all-time data what is apparent in figure 4 already, where data are split into season periods: delta(resilience) ≈ 0 for the whole premier soccer league Bundesliga. Within acceptable levels of confidence there is no resilience effect.

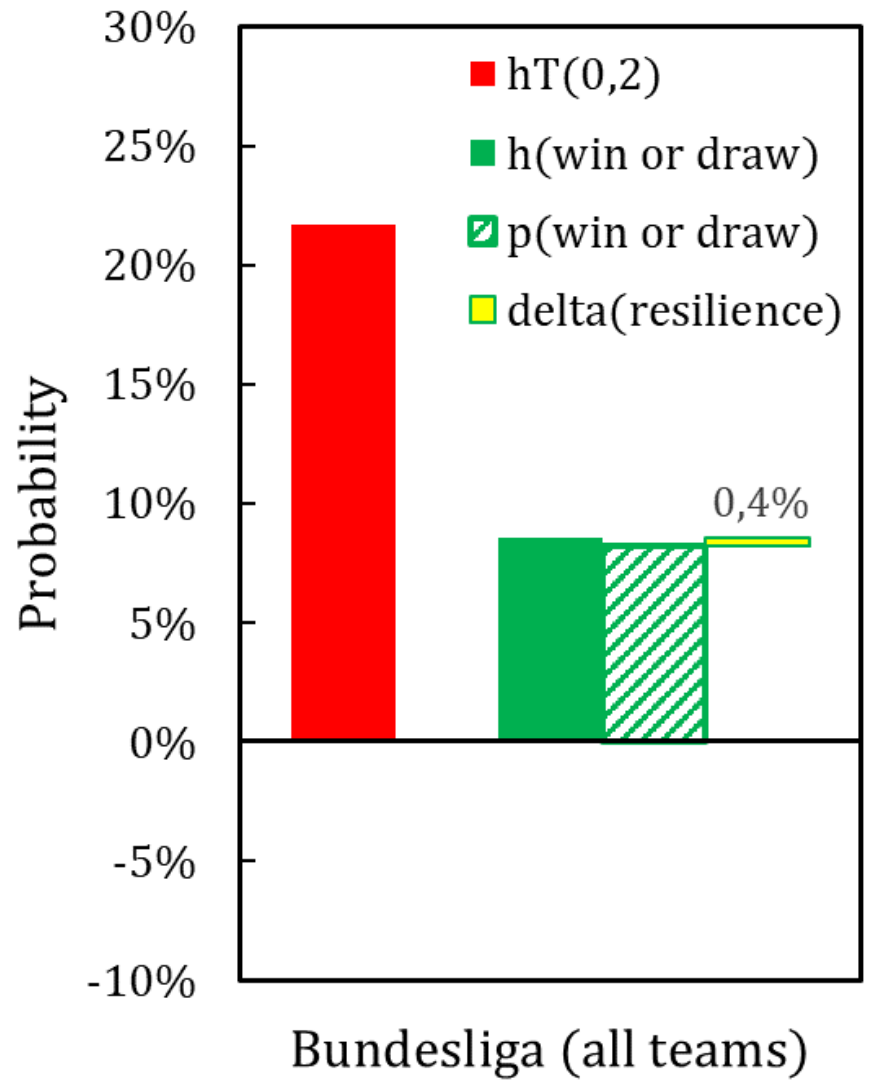

**Figure 5**: All-time relative frequencies hT(0,2) and h(win or draw) for the soccer league Bundesliga. The theoretical probability p(win or draw) is positioned alongside the empirical relative frequency h(win or draw). The difference between h(win or draw) and p(win or draw) is interpreted as resilience effect and provided as delta(resilience). For the whole league, the resilience effect does cancel out to nearly zero.

The outcome is different when the data are analyzed on soccer team level. As figures 6a and 6b illustrate, the top 10 teams do suffer less 0:2 leeway, the teams following on ranks 11 to 20 do suffer more 0:2 leeway, compared to the average value hT(0,2) = 21.7% for the whole league. The team FC Bayern München on the first rank of the all-time table displays the lowest value with hT(0,2) = 11.1%, followed by the team Borussia Dortmund on the second rank with hT(0,2) = 16.7%. hT(0,2) can be put into equation 6 to derive the goal scoring probability $p_A$ = 1 – SQRT[hT(0,2)], which for FC Bayern München yields $p_A$=67%, and for Borussia Dortmund $p_A$=59%. Assuming that $p_A$ remains constant after trailing by 2 goals, p(win or draw) is calculated according to equation 9, which for the team FC Bayern München yields p(win or draw) = 8.2%. This value contrasts h(win or draw) from empirical data, which is 21.1%. According to equation 10, the offset between h(win or draw) and p(win or draw) is attributed to resilience. With 12.9%, FC Bayern München exceeds all other teams. delta(resilience) can be comprised of two contributions. First of all, the probability to reach a draw or to win the match increases when $p_A$ increases after having conceded 2 goals. For FC Bayern München, $p_A$ has to increase from 67% up to 98% to explain the extraordinary high value of h(win or draw) = 21.1% for a 3:2 win or a 2:2 draw. Secondly, the team could manage to win a lot more matches with even higher scores than 3:2 when $p_A$ increases, which adds up to the right hand side in equation 8. Due to a lack of granularity of the data, these effects cannot be separated further. However, a sudden increase of team strength and hence an increase of $p_A$ is required for both contributions.

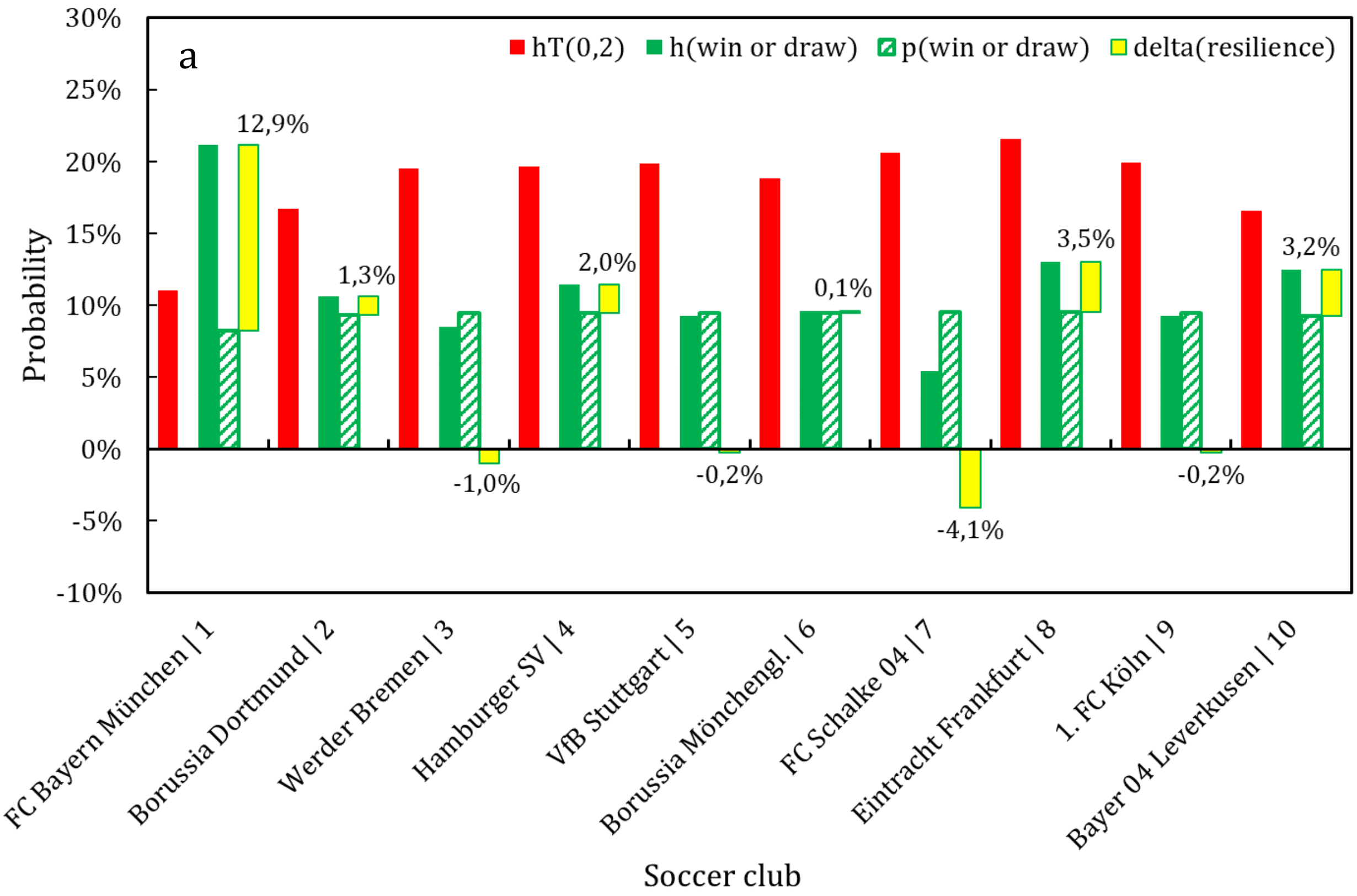

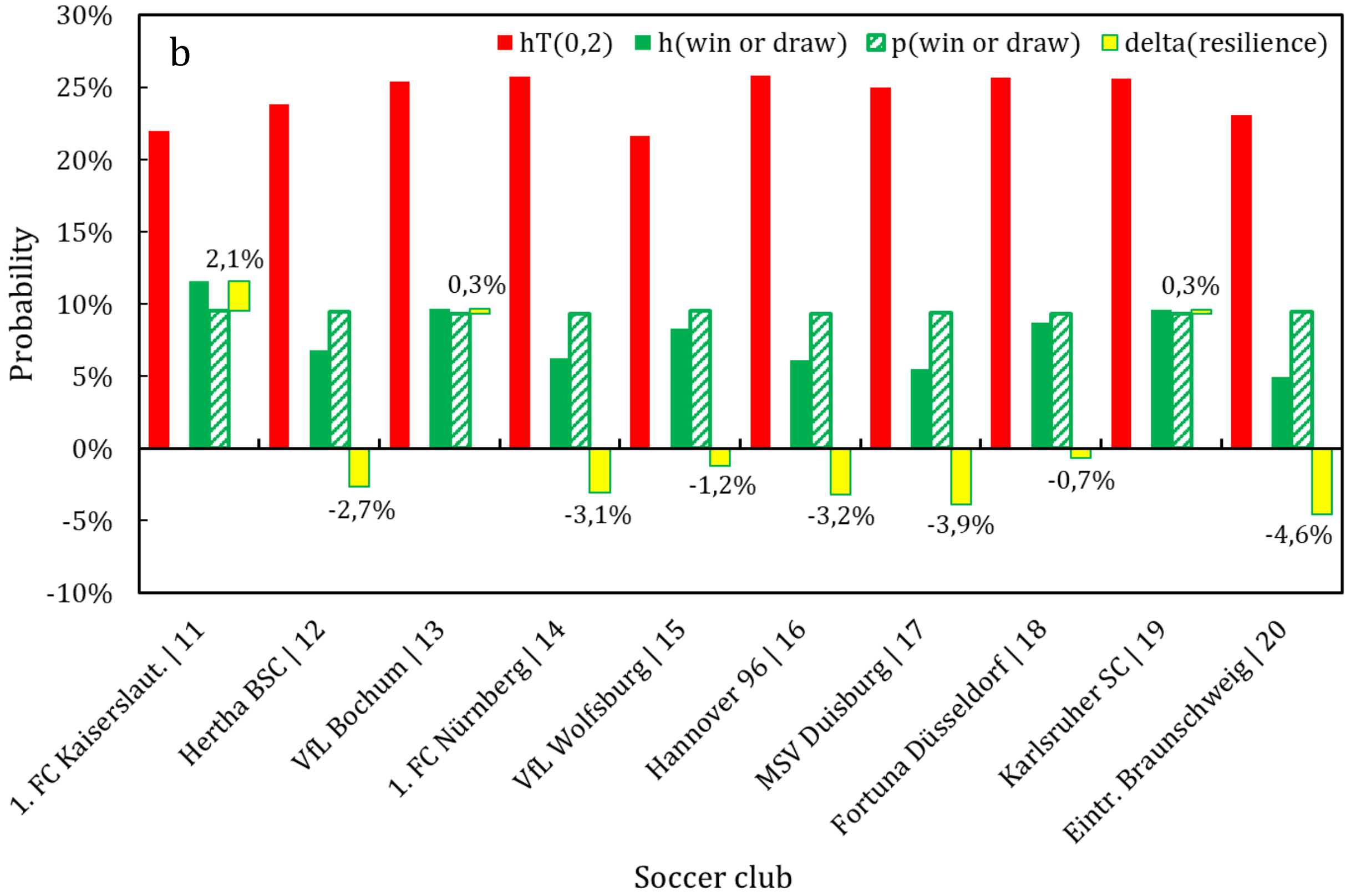

**Figures 6a and 6b**: All-time relative frequencies hT(0,2) and h(win or draw) of the first 20 clubs from the all-time table of the Bundesliga. The theoretical probabilities p(win or draw) are positioned alongside the empirical relative frequencies h(win or draw). The difference between h(win or draw) and p(win or draw) is interpreted as a resilience effect and provided as delta(resilience) for each team.

The relative frequency h(win or draw) of the majority of the top 10 teams is above 8.6% average, see filled green columns in figure 6a, except for two teams. With 8.5%, the club Werder Bremen is close to average, and with 5.4%, the club FC Schalke 04 is below average. As already mentioned, FC Bayern München displays an exceptionally high value: if the team trails by 2 goals, which happens with a relative frequency of hT(0,2) = 11.1% only, it turns these matches to its favor with a relative frequency of h(win or draw) = 21.1%, which is 2.5 times as much as teams in the Bundesliga are able to achieve on average. This empirical finding is extraordinary in the sense that the team following on the second rank, Borussia Dortmund, has a value of h(win or draw) = 10.7%, which is nearly half the value of FC Bayern München. Within acceptable levels of confidence, only Werder Bremen and FC Schalke 04 display a negative resilience among the top 10, which corresponds to a negative response to the initial setback of conceding 2 goals. In contrast to the top 10 in figure 6a, the majority of teams on ranks 11 to 20 in figure 6b exhibits lower performance. The relative frequencies h(win or draw) are below 8.6% average, except for the clubs 1. FC Kaiserslautern, VfL Bochum, and Karlsruher SC. The majority of teams exhibits negative resilience. Within acceptable levels of confidence only the 1. FC Kaiserslautern exhibits a positive resilience.

Overall, the distribution of resilience effects accumulates with rankings in the all-time table. The majority of teams with positive resilience is positioned among the top 10 clubs, the majority of teams with negative resilience, which relates to discouragement effects, is found on the lower ranks 11 to 20. However, a clear correlation between delta(resilience) and rank in the all-time table is lacking, except for the club on first rank, the FC Bayern München.

### 4.3. Resilience and the miracle of Bern

The analysis above allows to shed some light onto the miracle of Bern. More than 60 years have passed, and many studies have been published about the match and the events associated with it [21]. The soccer teams met on neutral ground, so home advantage effects are not present. Applying the statistics from the Bundesliga as a first guess, one can assume a probability of pT(0,2) = 21.8% for trailing by 2 goals. Changing the odds into scoring 3 goals in sequence and winning the match 3:2 has a probability of p(win) = 2.7%, which can be rightly described as a rare event.

In a next step, it is examined whether the probability of the outcome of the match truly represents the relative strengths of the two teams. In the group phase, Hungary achieved 17 goals and conceded 3 goals, the probability for Hungary to score a goal can be estimated as $p_{Hungary} \approx 85\%$. For Germany, the same consideration leads to a probability $p_{Germany} \approx 44\%$ (7 goals scored, 9 goals conceded). It appears that the Hungarian team was approximately 2 times stronger than the German team, which was confirmed by the outcome of the match in the group phase, where Hungary won with 8:3. In the aftermath, the score was excused with the strategic move of the German trainer to rest some key players, safe in the knowledge that loosing this match wouldn't affect their fate. Both goal scoring probabilities $p_{Hungary}$ and $p_{Germany}$ are assumed to remain constant throughout the match and feed into the Poisson model in equation 2. More than 60 years ago, the expectation value E for goals per match differed from todays values. For the FIFA world cup in 1954 with 140 goals scored in 26 matches, $E = 5.4$, which has to be taken into account. The probability for Germany to win a match versus Hungary calculates to 16% (all probabilities p(k,l) with $k > l$ added up), a draw to 14% (all probabilities p(k,l) with $k = l$ added up), and a loss to 70% (all probabilities p(k,l) with $k < l$ added up). Based on these figures, the probability for Hungary to win any match versus Germany was quite high and generally expected by all spectators. G. K. Skinner et al. [22] investigated the probability whether the final score of a soccer match truly represents the relative abilities of the competing teams. Applied to the 8:3 result of Hungary versus Germany in the group phase, the Bayesian approach therein yields a confidence level of 85.5% that the score truly represents the relative strengths of the teams. The value is close to the 90% confidence which is frequently considered a minimum acceptable level of confidence in the outcome of an experiment in quantitative disciplines. In contrast to that, applying the same Bayesian approach to the final match, the 3:2 win of the German team has a confidence level of 61.2% only. In other words, there is a probability of 38.8% that the 3:2 win does not correctly represent the true team strengths. It was argued later that the German team was better able to cope with rainy conditions present in the final match, equipped with studded soccer shoes which provided better grip on soaked ground.

Dispensing with the assumption of miracles, the 3:2 win in the final match can be explained under the condition that the goal scoring probability of the German team, hence its underlying team strength, has increased considerably. In order to increase p(win) in equation 8 onto a 2-digit level, the goal scoring probability of Germany would need to increase to $p_{Germany} \approx 80\%$, after trailing by 2 goals. A similar fundamental increase of goal scoring probability needs to be assumed to forecast a win for Germany applying the before mentioned Poisson model in equation 2. Within the scope of the present model, the increase of team strength after the 0:2 leeway can be interpreted as an indication for resilience. Summarizing all findings without assuming divine interference, the surprise in the final match in the 1954 FIFA world cup might have been due to a lucky combination of several factors. This includes tactics (positioning of key players), technology (studded soccer shoes) and resilience (strength of nerve).

## 5. Discussion and summary

This work has introduced a quantitative model for determining the resilience of soccer teams. The model is able to carve out from empirical data the two necessary conditions, (i) adversity and (ii) positive adaption to this adversity. Adversity is represented by an initial 0:2 leeway, which constitutes a considerable setback for the trailing team. Successive positive adaption is represented by turning a probable loss of the match into a win or at least a draw. Statistical data are compared with a theoretical model, making use of data from the German premier soccer league Bundesliga over the last 59 seasons, beginning at the first season 1963/64. It is summarized in figures 6a and 6b that the majority of the top 10 soccer clubs listed in the all-time table exhibit positive resilience, whereas the majority of teams on the following ranks 11 to 20 exhibit negative resilience, which can be related to discouragement effects. Within the framework of the present model, the soccer team FC Bayern München shows extraordinary high resilience. Comparing its empirical relative frequency h(win or draw) = 21.1% with its base value p(win or draw) = 8.2%, it is able to turn 2.6 times more matches than expected to its favor, when the team releases additional power and increases its goal scoring probability $p_A$ to 98% after trailing by 2 goals. FC Bayern München considerably differs from all other soccer teams of the league, its resilience poses a unique feature. The root causes for being able to activate additional resources should further be investigated.

Compared to the average for the whole soccer league, the positive offsets for the top soccer teams indicate that resilience is an important contribution for success. Resilience accumulates with the top 10 teams, but there is no systematic stepwise graduation of resilience to teams on lower ranks as one might expect, which leaves room for speculation on the root cause. The lack of graduation possibly implies that there is no dedicated focus in soccer training programs onto systematically fostering resilience. The fact that positive resilience values are present for the majority of the top 10 teams might be attributed to the circumstance that resilience is unknowingly enhanced as a side effect while optimizing individual soccer gamer excellence [23]. Fostering resilience is still challenging to be put into practice, without obvious parameters to optimize for athletes, coaches, and sport psychology practitioners. At competitions, it is easier to count goals, ball contacts, fouls, running speeds and passes with the camera equipment available. It is much harder to carve out indications and measures for experience, for the ability to perform under pressure, for positive responses to setbacks, and for strength of nerve, which are crucial to resilience. For instance in other sport domains, mountain climbers and long distance runners exhibit high resilience levels, which relates to physical condition, experience, and grit. It requires endurance and time to work on those factors, which go far beyond talent and technique. Instead of focusing onto isolated performance parameters, recent proposals highlight systemic approaches as a key to excellence and provide practical recommendations for sports, including facilitative environments, to foster resilience [24, 25].

For uncovering resilience effects, it was necessary to quantify the home advantage in empirical data first. As an additional learning, it is shown that the home advantage is slowly fading away from the German premier soccer league Bundesliga with a rate of -2% per decade. It is expected to vanish in the season period 2032/33 – 2041/42. The main reason is the professionalization of the soccer league, with players changing club affiliations, gradually loosing the emotional ties to their home turf.

The miracle of Bern is a prominent example for a 0:2 leeway and an unexpected 3:2 win, which took place in the 1954 FIFA world cup end game between Germany and Hungary. A lucky combination of factors might have contributed to the surprising result. The underlying team strength and hence the goal scoring probability of the German soccer team has improved considerably. Within the scope of the present model, the increase of team strength occurring after the 0:2 leeway can be interpreted as an indication for resilience.